**Synthesis of $ReN_3$ thin films by magnetron sputtering.**

G. Soto, H. Tiznado, W. de la Cruz, A. Reyes.

*Universidad Nacional Autónoma de México, Centro de Nanociencias y Nanotecnología*

*Km 107 carretera Tijuana-Ensenada, Ensenada Baja California, México.*

**Abstract**

Recently was reported a novel compound between rhenium and nitrogen, announced with $ReN_2$ composition. This compound was synthesized by the high temperature and high pressure method. We found that the diffraction peaks of this compound are in agreement with the x-ray pattern of a rhenium-nitrogen film, under the assumption that the film is oriented on the substrate. The film was prepared by reactive magnetron sputtering, at room temperature, and deposited on a silicon wafer. From the analysis of the diffractograms it could be concluded that both materials share the same structure. By density functional calculation was found that the composition could be $ReN_3$, instead of $ReN_2$, as stated before. The $ReN_3$ fits in the *Ama2* (40) orthorhombic space group, and by the existence of $N_3$ anions it should be categorized as an azide; that is, a nitrogen-rich compound. To reach high nitrogen concentrations by sputtering a crucial step is the target-poisoning. Under this regime of deposition is ensured that the compound is formed simultaneously on the substrate and the target. The poisoned target is rarely used because of a reduced sputtering yield, but as shall see, it can be used as a novel synthetic technique.

**Corresponding Author:** **G. Soto.**

**CNyN-UNAM**

**P.O. Box 439036, San Ysidro, CA**

**92143-9036, USA**

**Tel: +52+646+1744602, Fax: +52+646+1744603**

**E-Mail: gerardo@cnyn.unam.mx**

## 1. Introduction

In recent years has become common to find reports on the synthesis of novel materials by High-Pressure and High-Temperature (HPHT) methods. The use of extreme pressures and temperatures, combined with compositional variables, provides an opportunity to synthesize uncommon materials and/or to tune their physical properties for a wide range of applications [1]. HPHT methods were pioneered by researcher trying to synthesize diamonds, but are now often applied to find materials that might have mechanical advantages and chemical and thermal resistance comparable-with or superior-to those of the diamond. The transition metal nitrides are promising as hard materials and can be used for machining ferrous alloys. By HPHT were synthesized novel nitrides of heavy metals such as platinum, osmium and iridium [2, 3]. Due to the observed high bulk modulus of the above materials, some of the latter HPHT-studies are aimed to find novel nitrides of heavy metals. A example of this effort is the work of Kawamura *et al.* who recently published an article about the HPHT-synthesis of $ReN_2$ [4].

In direct competition with the HPHT methods for the synthesis of artificial diamond are the methods of plasma[5]. These have been successful in terms of quality and purity of the obtained diamond, with the added advantage that the materials can be applied as coatings directly from the synthesis. Similarly, the plasma methods are currently used to find alternative materials to replace the diamond. The success of plasmas to produce new materials originates from the high reactivity obtained by breaking the molecular bonds after multiple ionization events. By laser ablation experiments -a very energetic plasma method- was possible to growth films of $AuN_x$ [6], $PtN_x$[7], and $RuN_x$ [8]. In comparison with HPHT, the main limitations of the plasma synthesis are the low growth rates and the textured crystalline orientations, which makes difficult to fully characterize the materials obtained. Due to the inherent impossibility to determine unambiguously the crystalline arrangements, the research of heavy metal nitrides by plasma methods has not gained much relevance.

Recently we published an article about the synthesis of rhenium nitride by a plasma method [9]. We were successful in explaining the experiment when nitrogen is in low concentrations. Basically it is a solid solution of nitrogen incorporated into the octahedral interstices of the rhenium lattice. However we failed in interpreting the data when nitrogen is found in high concentrations. Recently Kawamura *et al.* reported the formation of a rhenium-nitrogen compound, assumed with $ReN_2$ composition and $MoS_2$-like structure [4]. The report of Kawamura was very useful to resolve the nitrogen-rich rhenium compound. Thereafter, the objective of this report is to present the kind of compound that result when a rhenium target is deposited by dc-sputtering in a heavily nitrogen-poisoned state. As a clarification, the target-poisoning means that the compound is not only formed on the substrate as desired, but also on the sputter target, which results in a significantly reduced sputter yield and, thereby, reduced deposition rate [10].

This regime of deposition is seldom used, but as we shall see, it can be useful to prepare a novel compounds.

**2. Our previous work.**

The experimental setup and method for film preparation are described in the appendix A, this is the same methodology in use since the previous work [11]. The nitrogen-rich rhenium film was prepared at room temperature and 180 mTorr of $N_2$ ($F_N$ = 35 mL min$^{-1}$, $F_N$ is the nitrogen flow), with an additional flow of Ar, $F_{Ar}$ = 5 mL min$^{-1}$ ($P_{Ar}$ ~ 5 mTorr). Fig 1 exhibit several XRD patterns of rhenium nitrides grown at different nitrogen flow rates. This figure is important to make an association from the low-nitrogen to the high-nitrogen contents. The interpretation of diffractograms is as follow.

The peaks indicated with the * symbol comes from the silicon wafer, used as substrate, which has peaks at 28.40° (111) and 58.85° (222). The sample deposited with $F_N$ = 0, Fig 1(a), shows three main diffraction peaks at 37.51°, 40.4° and 42.8°. These correspond to the diffraction planes (100), (002) and (101), respectively, of the hexagonal close packed structure reported in the Powder Diffraction File (PDF) #89-2935. The (002) peak correspond to a Re-Re interplanar spacing of 2.23 Å in *c*-direction. The pure-Re film is to some extent preferentially oriented, with (002)-Re parallel to (111)-Si. For the $F_N$ = 16 mL min$^{-1}$ sample, Fig 1(b), the hexagonal Re peaks were not present. Instead, wider peaks at approximately 38.45° (A), 44.69° (B), 65.20° (C) and 78.1° (D) appear in the pattern. This pattern agree with the face center cubic phase reported by Fuchigami *et al.* [5], with $a_o$ = 4.03± 0.03 Å. This pattern also agrees with the DFT computer-generated diffractogram (not show here) for *cubic*-$Re_4N$ with N in octahedral interstices. For higher values of $F_N$, the positions of peaks A through D shift to lower angles as the N/Re ratio in the films increase. The shift is more noticeable for the A-peak, corresponding to the (111)-planes when referenced to a cubic phase. In Fig 1(c) it shifts to 38.16° (Re-Re interplanar spacing of 2.35 Å) and then to 36.46° in Fig 1(d) (Re-Re interplanar spacing of 2.46 Å). These shifts indicate that the crystalline structure is expanding with increasing nitrogen content. For the most important diffraction peaks of Fig 1 we show the dependence of interplanar distance as a function of nitrogen flow in Fig 2. Additionally in Fig 1(d), for the $F_N$ = 30 mL min$^{-1}$ sample, two new broad peaks appear at approximately 51.02° (E) and 65.44° (F). These peaks cannot be indexed to a cubic phase, and thereafter are related to a different arrangement of rhenium nitride, given their shape and position. In the same sample, some peaks were present at 16.45° (110) and 25.47° (210), which might be indexed to a hexagonal crystalline phase of $ReO_3$ using PDF #40-1155.

At the time that this film was synthesized we conducted an exhaustive search, trying to index this pattern to a known phase of rhenium nitride. As we found, no phase in the PDF database or reviewed literature for rhenium in metallic, nitride or oxide, is in agreement with the diffraction pattern presented in Fig 1(d). However it is clear from Fig 2 that there is a noticeable expansion of the crystal lattice as a function of the nitrogen incorporation. The expansion occurs preferentially along the close-packed direction of rhenium atoms, that is, (002) when referred to the hexagonal structure of metallic-Re, and (111) when referred to cubic structure of $ReN_x$.

### 3. The Kawamura's experiment.

As mentioned above, Kawamura *et al.* reported the formation of a rhenium-nitrogen compound [4]. The report of Kawamura was very useful to resolve the nitrogen-rich rhenium compound. We would like to briefly summarize the results of Kawamura here. This article shows the HPHT metathesis reaction between lithium nitride and rhenium pentachloride which produce materials with platelet shapes. Kawamura report powder x-ray diffraction (XRD) patterns using Bragg-Brentano and Gandolfi geometries. All diffraction peaks where indexed by Rietveld structure analysis to a hexagonal *P63/mmc* (194) space group, isostructural with $MoS_2$ and similar to $ReB_2$. The Rietveld analysis gave lattice constant of $a$ = 2.806 Å and $c$ = 10.747Å, and V = 75.28 $Å^3$, and an atomic coordinate of N($z$) = 0.616. XPS analysis where used to confirm the Re-N bonding. The authors conclude that their synthesis gives $ReN_2$. The structure parameters differ with some theoretical calculations of rhenium-nitrogen compounds in 1:2 stoichiometric ratios [12–14]. While the reason for the discrepancy were not clarified, it was assumed that this type of synthesis often develop unusual metastable phases. The unit cell and structural parameters as are proposed by Kawamura are shown in Table I.

### 4. Comparison of sputtered XRD pattern and Kawamura's proposed structure.

In Fig 3 are compared the experimental diffraction pattern of the film growth at $F_N$ = 30 mL $min^{-1}$ with the generated diffraction pattern as of Kawamura's structure description. Fig 3(a) is the experimental thin-film pattern using the Bragg-Brentano geometry for the Cu-$K_\alpha$ radiation. Fig 3(b) is the generated powder diffractogram for the structure as it was proposed by Kawamura using the Cu-$K_\alpha$ wavelength. In Fig 3(c) are the best-match peaks from the powder pattern, which are associated to the experimental pattern. In Fig 3(d) is the diffractogram for the Kawamura's structure, in the assumption that this is a highly oriented material in the $c$-direction using Bragg-Brentano geometry; this is the pattern that Kawamura reported in the inset of Fig 3 in ref [4]. As it can be seen from this association, the XRD pattern of sputtered film could be indexed to the material synthesized by Kawamura *et al* [4]. However, the film that was grown in the sputtering experiment has a textured orientation which is different from that seen by Kawamura. As it is know, the texture is the distribution of crystallographic orientations of a polycrystalline sample. In the film the crystallographic orientations are not random, but have preferred orientations of (010)-planes parallel to (111)-planes of the Si substrate. There are some contribution from

(014) planes, indicative of some tilt in *c*-direction, and some contribution from (120) and (124). Fig 4 is a sketch where these possibilities are illustrated. In conclusion, the (010)-planes of the compound are parallel to the (111)-Si orientation, with slightly misoriented grains around the preferential orientation. As a consequence of texture, there are missing reflections from the sample in the Bragg-Brentano geometry, and to some extend weak count rates. In contrast, the material synthesized by Kawamura is highly texturized in *c*-direction. We conclude it is very likely that the material that was synthesized by high pressure and high temperature has the same crystalline system than the material that was synthesized by the sputtering method.

## 5. Density Function Theory (DFT) calculations.

At this stage it might say that both materials share the same crystallographic system. To understand more deeply the nature of this compound, we performed DFT calculations using the structure as it was proposed by Kawamura as the initial point. The details of calculations are given in the B-appendix. The plot of total energy versus volume is show in Fig 5. The $dE/dv$ at the volume given by the structural parameters reported by Kawamura is not zero. The $dE/dv \neq 0$ means that there is no internal cancelation of forces; surprisingly this can not be a metastable point. This detail implies that the structure proposed by Kawamura is wrong in some way. This becomes evident since the energy of $ReN_2$ at the point of Kawamura is high in comparison with the energy of the structure relaxed by DFT. In addition, the volume at the point of Kawamura is larger than the volume of the relaxed structure. It is unlikely that a structure that was synthesized under pressure develop a metastable phase that is a “tensioned variation” of the relaxed structure. Very recently it was proposed, using DFT structure optimizations, that the $MoS_2$-like $ReN_2$ structure will reform due to an rearrangement of the $N_2$-dimers, in such way that the hexagonal symmetry is loss [15]. Evidently, the diffraction pattern of the DFT-improved structure will not match the x-ray diffraction reported by Kawamura and thereafter, it will not match either the diffractograms obtained in this work. Additionally, in reference [15] it is found the $MoS_2$-like $ReN_2$ structure to be mechanically unstable according to the Born-Huang criterion for a hexagonal structure, since a negative value for the $C_{44}$ elastic constant was found. A negative $C_{44}$ means that the structure is tensioned, and the related strain tensor needs to be relaxed in the opposite direction. That is the same that we found and is illustrated in Fig 5, where the relaxation of $c/a$ to lower ratios is a crucial step to find the lowest-energy point. In brief, the structure as it was proposed by Kawamura lacks something along the *c*-direction. The DFT calculations reveal that the empty space in *c*-direction can not be justified by chemical bonds or Van der Waals forces. However, the Kawamura’s structure is correct in the sense that reproduces the experimental results, but at the same time is incorrect, because can not be a metastable phase. With the aim to elucidate this inconsistency, it was conjectured that there is plenty of space between Re-layers to insert the $(N_3)^-$ azide anions instead of $(N_2)^{2-}$ diazenide anios. The azide functional group is larger than the diazenide group; the larger anion could match the interplanar spacing between Re-layers. Table I list the structural parameters resulting from DFT optimization of $ReN_3$ constrained within the *P63/mmc* (194) space group. As it can be seen, the structure relaxed by DFT in the $ReN_3$ composition is a better match for the Kawamura's cell parameters. This gives the idea that the compound may be $ReN_3$ rather than $ReN_2$. However the nitrogen and rhenium atomic positions are highly constrained when the *P63/mmc* (194) space group is used. To avoid this, the structural degrees of freedom were increased by reducing the symmetry of the space group. From the initial *P63/mmc* (194) it goes down to the *Am2a* (40) space group. Even in the low-symmetry *Am2a* space group the diffraction pattern is closed the Kawamura's pattern, as can be seen in Fig 6. More DFT refinements are recommended to get a better match.

## 6. **Nanoindentation of $ReN_3$.**

Fig 7 shows the load-unload displacement cycles with limits of applied load of 50, 500 and 1000 μN for the $ReN_3$ sample. The cycles consist of 5 s of an increasing ramp up to the maximum load, 5 s of sustaining the maximum load, 5 s of discharge, and 3 s of added acquisition at the contact force. As seen, this film shows a very peculiar behavior. For a typically sample in the ramp-up, when the indenter contacts the material, is observed that the sample initially undergoes elastic deformation until the applied load reaches a critical value. Exceeding this critical value and the plastic deformation takes place, which can be characterized by sudden depth incursion or "pop-in" [16]. The peculiarity of this material is the number of pop-ins in the load-displacement curve. As we can see from inset of Fig 7, the elastic deformation is extremely small before the material experience plastic deformations. The sudden insertions appear in the plot as vibrations of the Berkovich tip. The deformation is inherent unstable, occurring in burst of highly localized strains [17]. There are also large pop-in events that can be observed in the load–displacement curves, $\delta_{exc}$. Taken as a whole, the deformation is sum of small and large pop-in events that start to happen at infinitesimal loads. When a material have the tendency to fracture with very little or no detectable deformation beforehand is considered brittle. In such case we can consider this material to have a perfect-brittle mechanical behavior, there is not elastic deformation. When the load is held constant the plastic flow is evident because the penetration depth increases noticeably. For a film with thickness of about one micron (estimated by means of x-ray measurements), a creep of 100 nm is outstanding. An analysis of the initial portion of the unloading response gives an estimate of the elastic modulus of the indented material. As we can see, the initial slope of the unloading portion approach zero, $\frac{dP}{dH} \to 0$, which means that this material is very soft. A detailed analysis of the raw data (hardly visible at this plot-scale) shows extremely small "pop-out" events in the unloading portion of the curve. The pop-outs are attributed to the release of internal stresses generated during loading [18]. Some the deformation is recovered by the pop-outs. The strangest thing is that this material fully recovers of its deformation after removal of the load. To be exact, the hysteresis curve returns to the initial point upon removing the load, as seen in Fig 7. By using the Berkovich-tip as the AFM probe, it can be see that it has no vestige of indentations, as shown in Fig 8 before and after indentation. We have not explanation for this behavior since, as previously stated, this is a brittle material. However, we dare to speculate that it have more to do with chemical changes (oxidation) in the sample than with a recovery of $ReN_3$ mechanical deformations.

7. **Discussion.**

As of the experiment, where a rhenium target was sputtered in nitrogen environments at different pressures, it can be concluded that the incorporation of nitrogen causes the expansion of the rhenium lattice, as it is seen in Fig 2. From the x-ray patterns of Fig 1 can be seen that the planes in *c*-direction (referenced to the *hcp* structure of Re) are strongly affected. Therefore, it is necessary to explain how the separation of rhenium atoms became so large. From our previous work [9] we know that single-nitrogen atoms in interstitial sites can not explain this interplanar separation, whether we consider octahedral, tetrahedral, prismatic or antiprismatic sites. The logical assumption is to consider the double occupation of interstitial sites. The twofold occupancy entails the probable occurrence of N to N interactions by means of the $(N_2)^{2-}$ diazenide anion. By DFT calculation was found that even with the $N_2$ anion, there are leftover spaces that cannot be justified by any kind of interatomic forces, whether if Van der Waals forces or chemical bonds are considered. It is also possible to consider threefold occupation of interstitial sites. With the $(N_3)^-$ azide anion in the DFT calculations was possible to get a better match with the x-ray diffractograms. The azide anion fits perfectly in the space between the Re-layers, and finally gives a justification for the expansion of the crystal lattice of rhenium. By the DFT calculations we get a feasible structure that could explain the diffractograms obtained from the sputtered thin film. This structure is also close to the structural parameters proposed by Kawamura *et al.* [4], as seen in Table I. The diffractograms of the DFT-generated rhenium azide and the Kawamura's structure are compared in Fig 6 and are, to some extent, similar. Virtually the Re-atomic positions are preserved, but the N positions are amended to accommodate new N-atoms. We believe that this structure can be improved to achieve a better match with the experiment. Still, the fit with the experimental diffractogram is relatively good. This result suggests that the material synthesized by sputtering, and also by HPHT, is a kind of rhenium azide.

Regardless of the relatively good agreement between theory and experiment, both the $ReN_3$ and the $ReN_2$ are positive enthalpy compounds. Recently was also reported the synthesis of $Re_3N$ by Friedrich [19] by HPHT. The energy difference between either of $ReN_3$ or $ReN_2$ with $Re_3N$ is about 300 kJ mol$^{-1}$, favorable to the formation of $Re_3N$. The energy calculations are being considered for publication elsewhere. This leads to the following question: Being the $ReN_3$ a compound of relatively low stability, how the sputtering method led to the synthesis of this substance? This is an interesting topic we'd like to talk a bit here. As is well known, the reactive dc-magnetron sputtering is used typically for making protective and functional coatings. Less explored is its use as a tool for the synthesis of new materials. However this technique combines several incidents that are beneficial for materials synthesis [20]. The magnetic containment of the magnetron field keeps the electrons orbiting in a semi-closed path that maximizes the probability of collision between atoms and electrons, increasing the fraction of ionized species. For instance, the multiple $e^-$-collisions with the background gas splits the $N_2$-molecule, which result in very reactive $N^*$ excited species. By means of ionization is inserted chemical reactive functionality onto the otherwise non-reactive species [21]. These excited species are able to react with pure rhenium and even with the rhenium nitride previously formed by the poisoning of the target surface. The excess of energy of 300 kJ $mol^{-1}$ required for the formation of $ReN_3$ are taken form the energy released by the de-excitation of $N^*$. It is assumed that the last stage of the reaction happens on the surface of the target, but more work is needed to confirm this. What we are saying here is that even the nitrides can be further nitrided by the $N^*$ species, which result in the formation of over-stoichiometric compounds. Another feature is the transfer of momentum that occurs among the eroded and gas-background species. The kinetic energy of the species arriving to the substrate surface is a very important aspect which determines the type of material that could be formed. If the kinetic energy is too high eliminates any possibility of forming compounds of marginal stability by the action of re-sputtering. The kinetic energy is limited by the gas phase collisions, thereafter can be controlled by means of the target-substrate distance and the background gas pressure. In this case the background pressure was very high for a sputtering experiment, $P_T$ ~185 mTorr, and the target-substrate distance roughly 10 cm. As a consequence, the chances of decomposing the material already formed by re-sputtering were decreased. Another important aspect is the growing temperature. Low temperatures are required if compounds of marginal stability are sought. In the magnetron sputtering the substrate temperature can be intentionally kept low by the use of low-powers during the sputtering experiment. In our setup the fortuitous combination of the above factors led to the formation of $ReN_3$. In summary, the synthesis of $ReN_3$ was possible because of the overabundance of $N^*$ ionized species, the pre-formation of $Re_3N$ in the target surface, the rapid thermalization of plasma due to the high pressure, and the low temperature during deposition.

The deposition regime where the target is heavily poisoned is usually avoided because it gives materials poor physical characteristics [20]. High roughness, porosity, low-density and mechanical instability of the nitrides grown in the poisoned regime is frequently reported, and attributed to the physisorption of $N_2$ molecules in the films. In physisorption the forces involved are intermolecular forces (van der Waals forces). By X-ray photoelectron spectroscopy we don't find any evidence of physisorbed nitrogen, with $N_{1s}$ binding energy of approximately 404 eV. The binding energies of $N_{1s}$ were 397.1 ± 0.1 eV, indicative that nitrogen is chemisorbed. However by the chemisorption of weak chemical groups the same "poor" physical features can be obtained, as seen in Fig 7. The difference is that in the chemisorption the forces involved are valence forces of the same kind as those operating in the formation of chemical compounds. How important this finding could be? The azides are sensitive explosives, and when subjected to a suitable stimulus, such as impact, heat, friction or shock, will explode. This is because the azide group retains their molecular character until a sufficient stimulus is applied to cause exothermic dissociation. The azide anion is inherently stable, but becomes unstable when subjected to small perturbations. This group of materials is important for several applications, including industrial and military. To our knowledge, there are no previous reports of the synthesis of heavy metal azides, except by the work of Kawamura [4]. Therefore, is believed that the search of such compounds by unconventional chemical routes, including the high-pressure high-temperature and the plasma methods, is a fruitful field. It would be worthwhile to explore this hypothesis.

## 8. Conclusions.

In summary, the diazenide group inserted in the rhenium interstices could not explain the diffractions patterns of the recently prepared rhenium-nitrogen compounds. The azide group in the rhenium interstices gives a better match for the diffraction patterns. By means of a heavily nitrogen poisoned rhenium target we were able to deposit a polycrystalline $ReN_3$ thin film. This material shares the same crystallographic arrangement that the material produced by the HPHT method [4]. The diffractograms given by DFT agree with the experimental evidence. The $ReN_3$ is mechanical unstable under indentation. The possibility of formation of metal azides by sputtering deposition is field that must be studied to avoid the implicit danger of azides.

### Acknowledgments

Thanks to D. Dominguez for technical assistance. Financial support from CONACYT grant #82984 is acknowledged.

**Appendix A. – Experimental methods.**

Rhenium nitride films were grown by reactive DC-sputtering. The experiment was carried out in a custom made ultra-high vacuum (UHV) deposition system. The system consists of three stainless steel chambers: sample loading, deposition and analysis. The analysis chamber is a PHI-548 apparatus equipped with X-ray photoelectron (XPS) and Auger electron (AES) spectroscopies, and it operates at $10^{-10}$ Torr by means of an ion pump. The sample loading and deposition chambers are independently evacuated to $10^{-8}$ Torr or better by a 130 L $s^{-1}$ turbomolecular pump. A 2.5 cm diameter Re target (99.9 %) was positioned in a magnetron gun, 10 cm above a (111)-silicon substrate. A mechanized rod is used to transfer the sample between chambers, which are isolated by gate valves. With this setup, the deposited films can be analyzed without exposure to the outside atmosphere. The rhenium nitride films were deposited at room temperature at different $N_2$ pressures in the following manner. In the deposition chamber, a mass-flow controller fed $N_2$ (ultra-high purity grade) at increasing flows to set pressures ranging from 3 mTorr (1 mL $min^{-1}$) to 180 mTorr (35 mL $min^{-1}$). After setting the $N_2$ pressure, an additional 5 mL $min^{-1}$ of Ar (ultra-high purity grade) were allowed into the chamber. The samples were grown at 120 watts for 50 minutes and were then immediately transferred to the analysis chamber. A reference sample, metallic Re, was deposited in the absence of $N_2$. The assessment of nitride formation was performed by comparative core-level energy shifts in the XPS in relation to the metallic Re sample. The XPS measurements used an Al K-α ($h\nu$ = 1486.6 eV) x-ray non-monochromatic line. The *x*-ray diffraction (XRD) patterns were collected ex-situ in a Philips X'pert diffractometer using the Cu K$\alpha$ line with $\lambda$ = 0.154 nm.

## Appendix B. Detail of density functional calculations.

The calculations were realized within the framework of the density functional theory. We employed the full potential linearized augmented plane wave method, as implemented in the Wien2k code [22] using the generalized gradient approximation [23]. We used muffin-tin radii ($r_{mt}$) of 2.0 au for Re atoms, and 1.0 au for N, and angular momenta up to $l$ = 10. We used 64 *k*-points in the irreducible part of the Brillouin zone. Optimization of the volume, $c/a$ ($b$ = cte), $b/a$ ($c$ = cte) and internal atom positions were performed recursively. Having arrived at the minimum energy structures, their volumes were changed around the equilibrium value, and their bulk moduli obtained after fitting the Murnaghan equation of state to the calculated volume-energy data. The compound formation energy from molecular nitrogen is computed as a difference between the ab initio total energies of $ReN_3$ and those of Re metal and molecular $N_2$ [24].

$$E'_{ReN3} = \frac{E_i^{Tot} - nE_{Me-bulk}^{Tot} - m\frac{1}{2}E_{N2}^{tot}}{m+n} \qquad \text{(A.1)}$$

Here *n* and *m* are the number of metal and nitrogen atoms, respectively. From that definition, less positive values mean higher stability toward spontaneous dissociation to Me and $N_2$ constituents. Similarly, the formation energy from free atomic nitrogen is computed as:

$$E''_{ReN3} = \frac{E_i^{Tot} - nE_{Me-bulk}^{Tot} - mE_N^{tot}}{m+n} \qquad \text{(A. 2)}$$

Since the molecular and atomic species cannot be treated directly within WIEN2k, the $N_2$ dimer was simulated by constructing a sufficiently large tetragonal unit cell (space group *P4/mmm*, *a* = 4.50 Å, *c* = 5.60 Å), locating $N_2$ in the vertices along the z-axis (N-positions (0, 0, ±0.0981)) and performing the calculation for a single *k*-point at the origin of the first Brillouin zone. The nitrogen atomic specie was carried with analogous calculation for a nitrogen atom placed into fcc unit cell (*Fm3m*, *a* = 15 Å). In order to check how well our calculations can be used for nitrogen-rich compound, we did equivalent calculations for $AgN_3$. We have minimized the forces acting on all atoms. Our calculations agree with the experimental parameters for $AgN_3$ and with previous calculations for that compound [25].

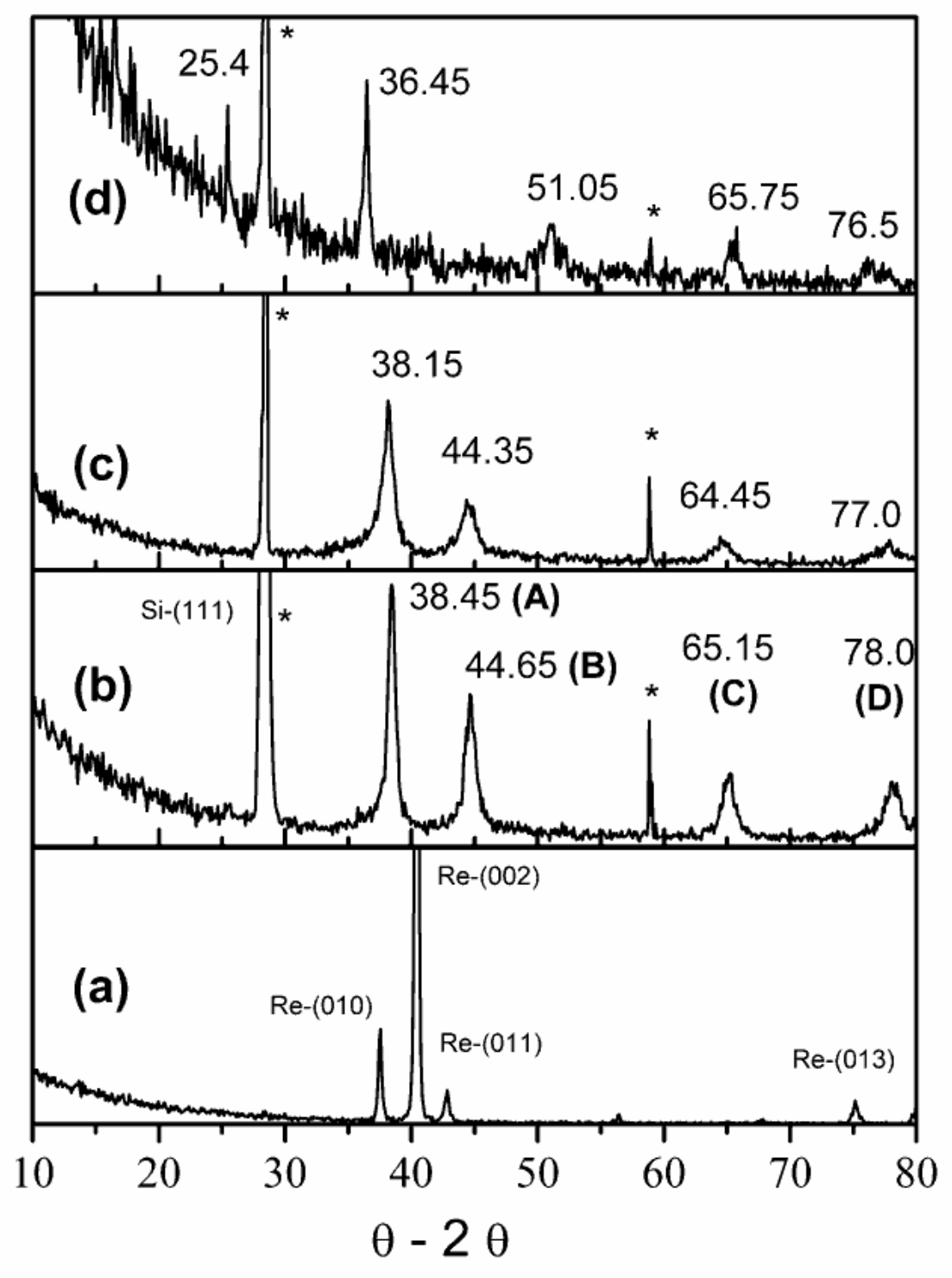

Figure 1. - XRD patterns for dc-sputtered thin films: (**a**) $F_N = 0$; (**b**) $F_N = 16$; (**c**) $F_N = 20$; (**d**) $F_N = 30$ mL $min^{-1}$.

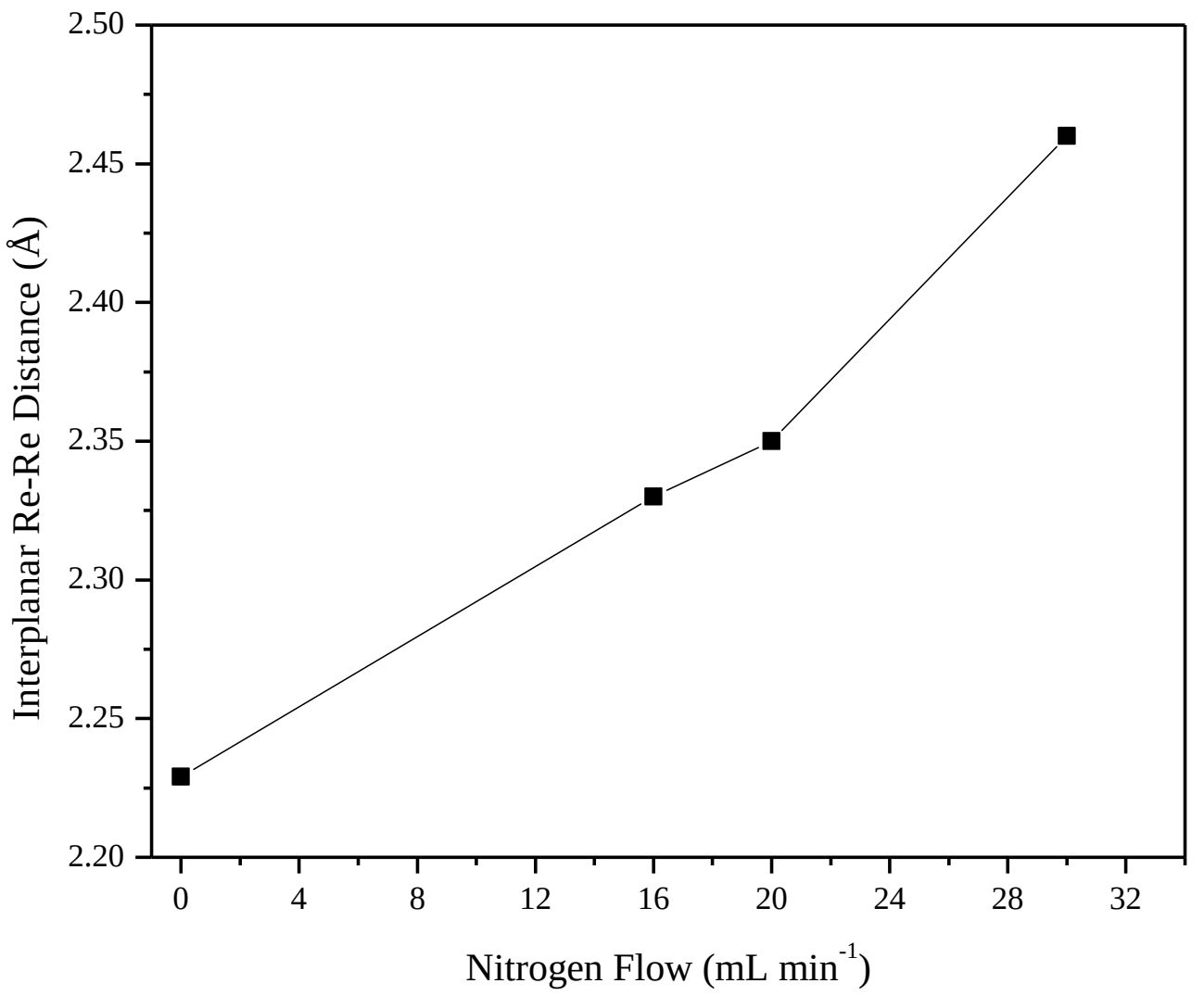

Figure 2.- Interplanar Re-Re distance as a function of nitrogen flow.

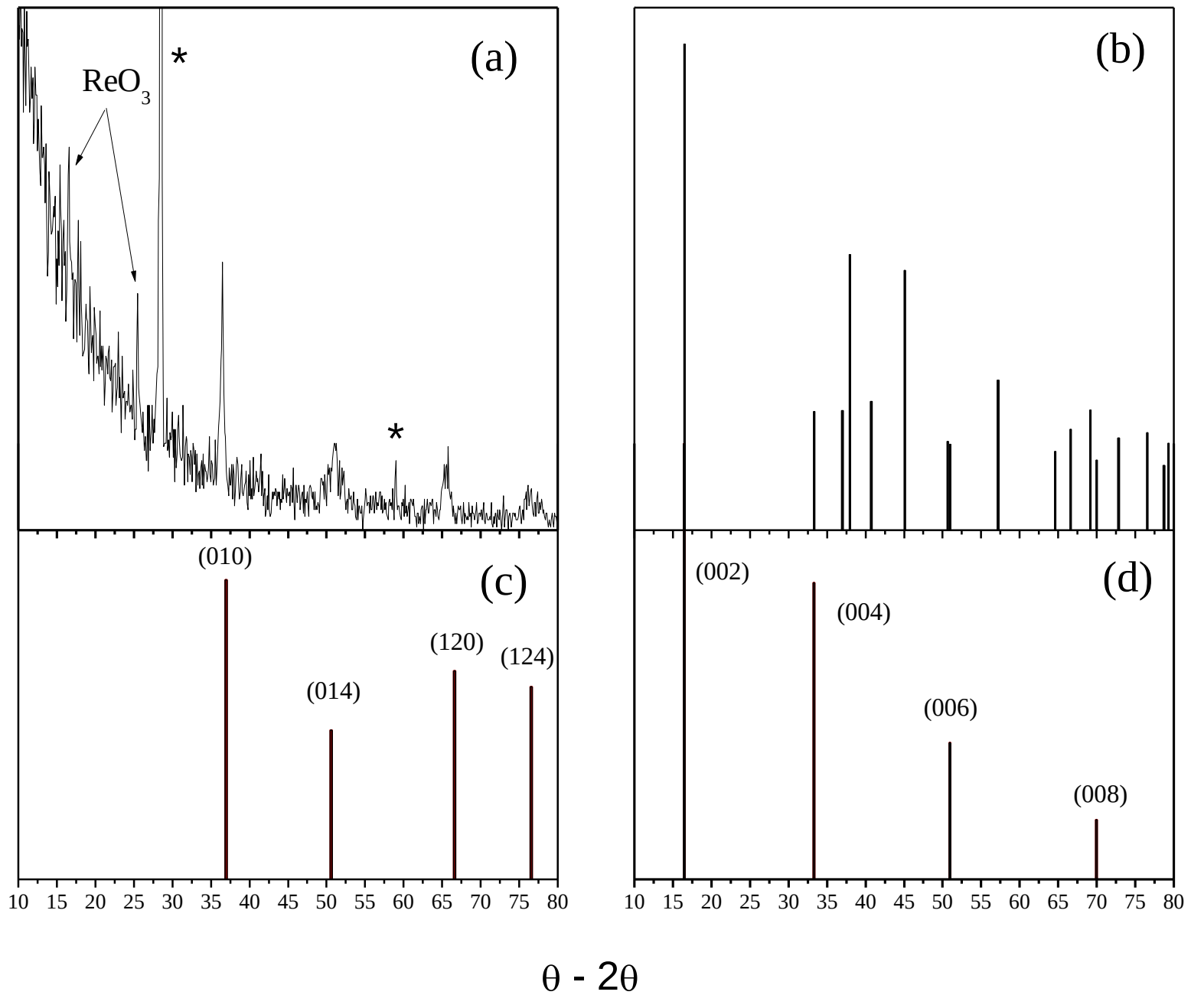

Figure 3.- Comparison of the diffraction patterns for (**a**) experimental $ReN_x$ film growth at $P_N$ = 30 ml min$^{-1}$, (**b**) powder pattern as of Kawamura data, (**c**) diffractions from Kawamura's data that match the experimental pattern, and (**c**) the pattern reported by Kawamura using Braggs-Brentano geometry.

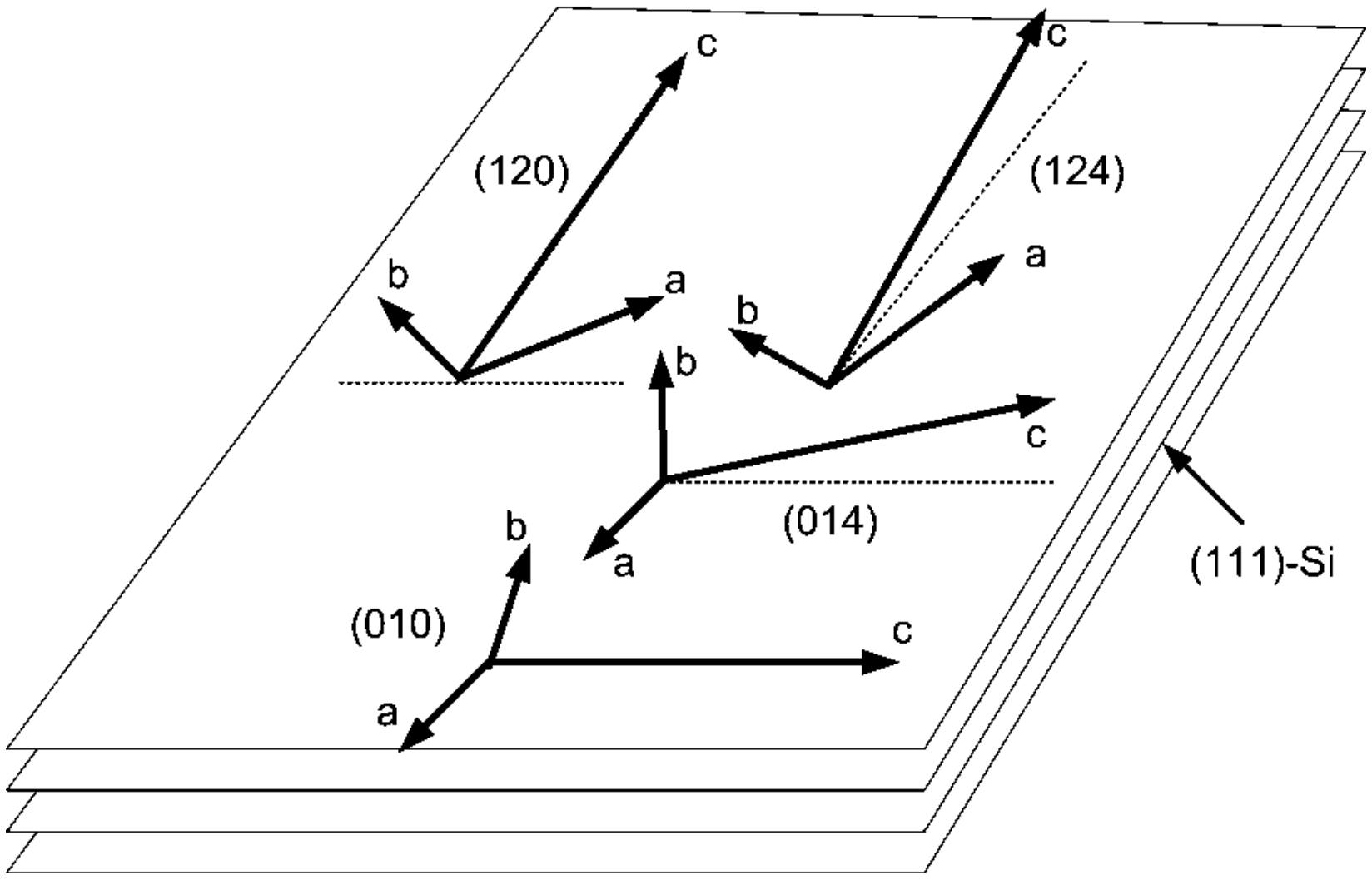

Figure 4.- Sketch of possible crystalline orientation of a textured films that might give the diffraction pattern of the experiment assuming the same structure as Kawamura.

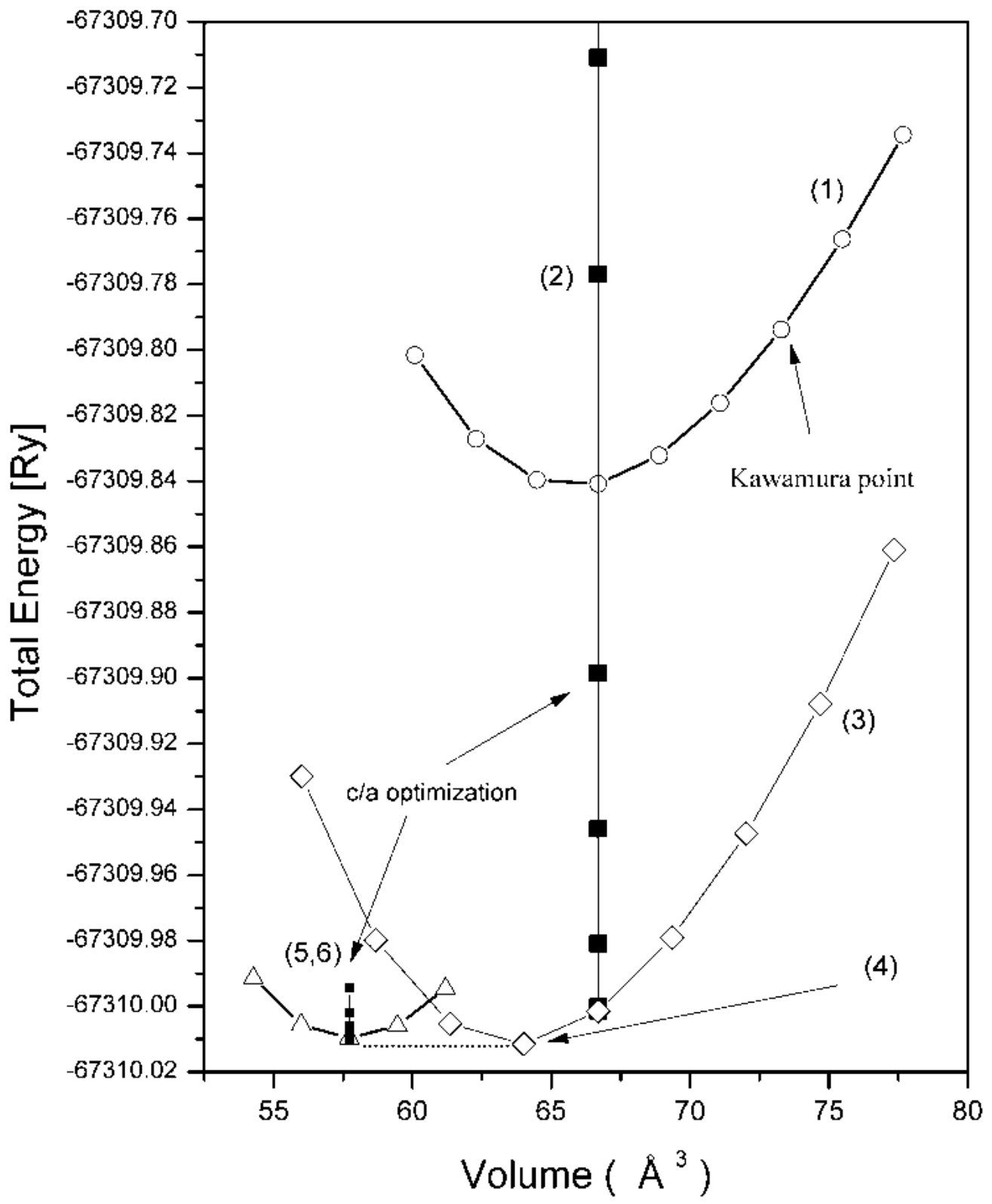

Figure 5.- Trajectory of structure optimization of $ReN_2$ by DFT.

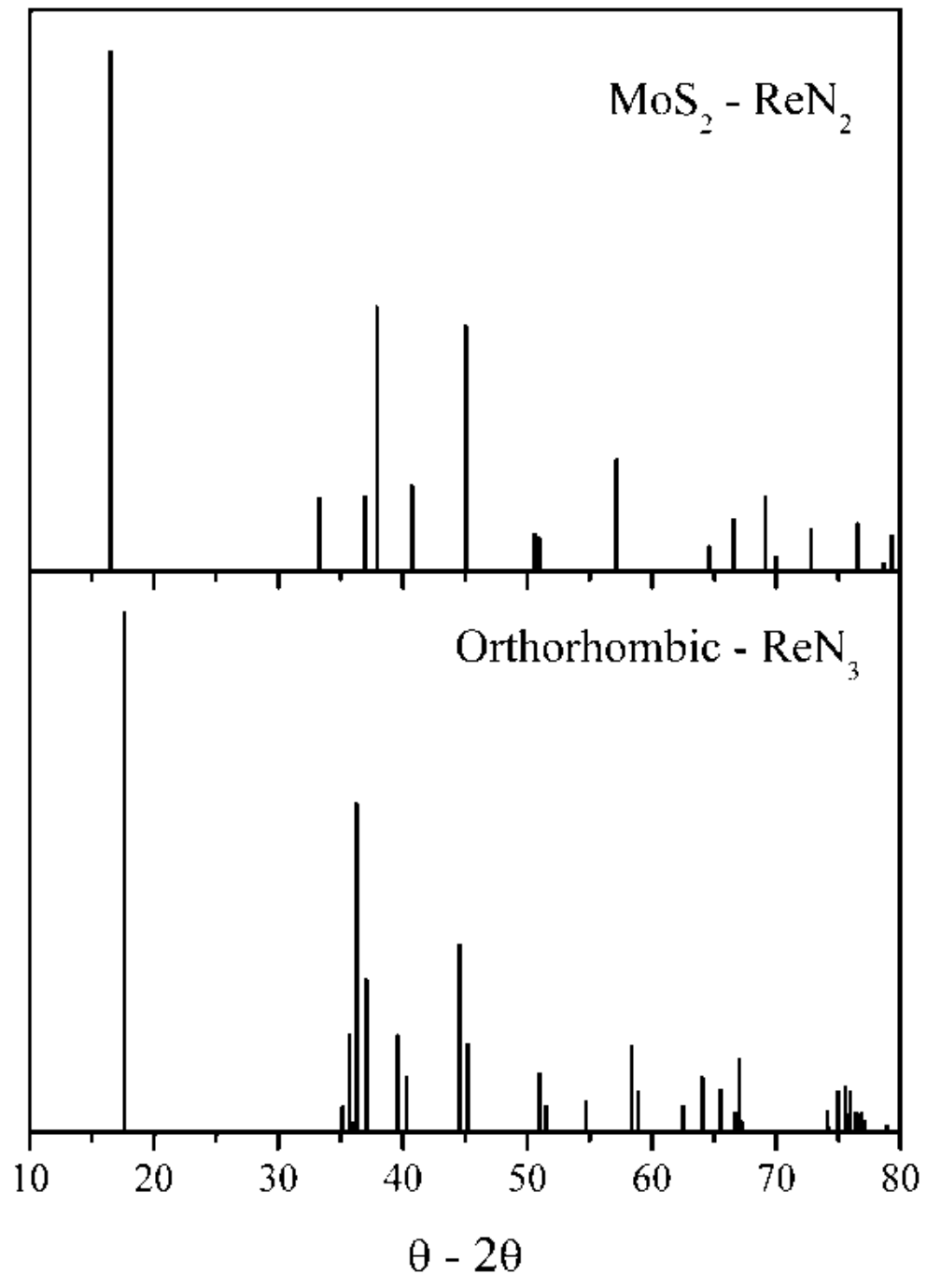

Figure 6. - Comparison of powder diffraction pattern of $ReN_2$ and $ReN_3$ using the Cu-kα radiation.

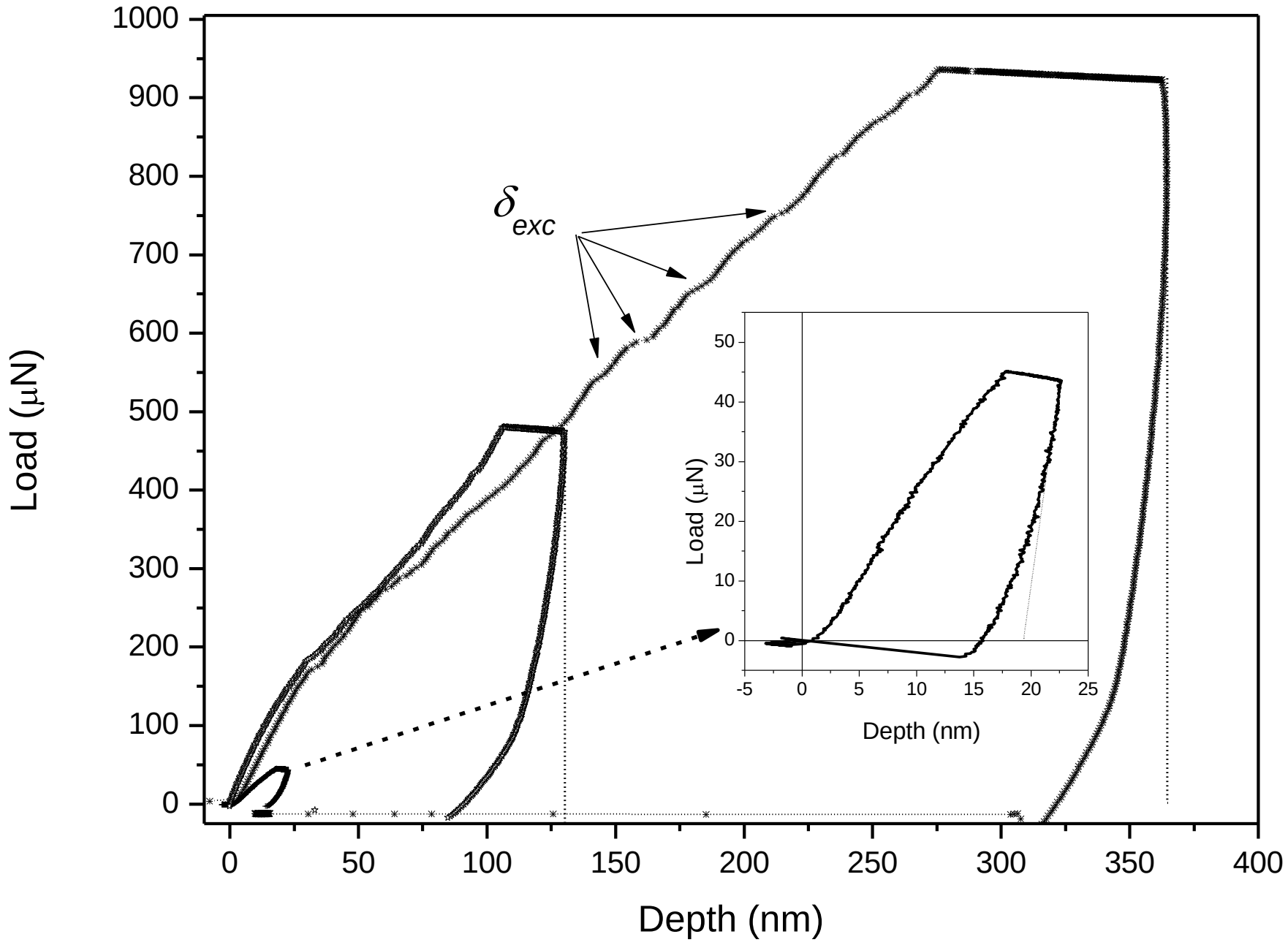

Figure 7.- Three different load-displacement curves with a Berkovich diamond nanoindenter for the sputtered film.

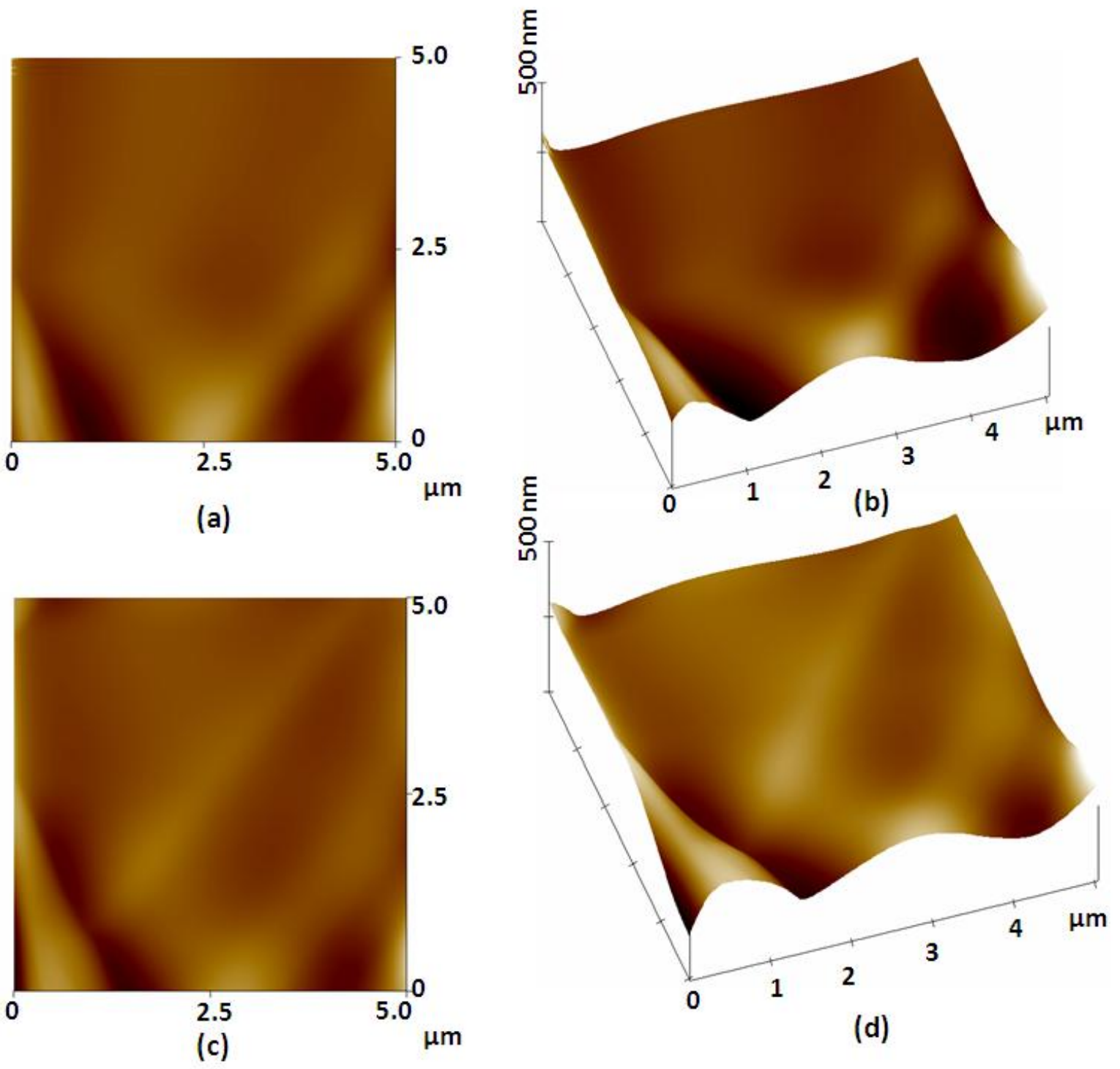

Figure 8.- AFM image taken before (*a* and *b*) and after indentation (*c* and d). Even being $ReN_3$ a brittle material, there are no traces of the indentation in the surface.

Table I. Compound, structural parameters and unit cells and DFT results of $ReN_2$ and $ReN_3$.

| Compound | Figure | Structural Parameters | Equation of state fit parameters[1] |
|---|---|---|---|
| $ReN_2$ | 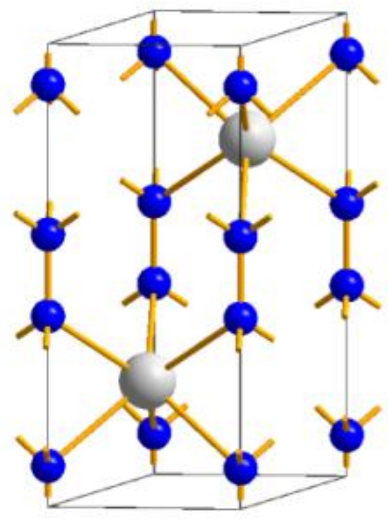 | By Kawamura *et. al.* (exp.)<br>Space Group: $P6_3/mmc$ (194)<br><br>$a_0 = b_0 = 2.81$ Å,<br>$c_0 = 10.75$ Å<br>$\alpha = \beta = 90$; $\gamma = 120$<br>Atomic Positions:<br>Re Wyckoff 2*d*<br>(0.66667, 0.33333, 0.25)<br>N Wyckoff 4*e*<br>(0, 0, 0.616) | $V_0 = 75.28$ Å$^3$<br><br>$B_0 = 173$ GPa<br><br>$E_t = -67309.793821$ Ry<br>(calculated by DFT)<br><br>$E_A = +0.55588$ eV atom$^{-1}$<br><br>$E_B = -5.86597$eV atom$^{-1}$ |
| | | Relaxed by DFT:<br>Space Group: $P6_3/mmc$ (194)<br><br>$a_0 = b_0 = 2.93$ Å,<br>$c_0 = 7.74$ Å<br>$\alpha = \beta = 90$; $\gamma = 120$<br>Atomic Positions:<br>Re Wyckoff 2*d*<br>(0.66667, 0.33333, 0.25)<br>N Wyckoff 4*e*<br>(0, 0, 0.088355) | $V_0 = 57.70$ Å$^3$<br><br>$B_0 = 350$ GPa<br><br>$dB_0 = 7.0710$<br><br>$E_t = -67310.00987$ Ry<br><br>$E_A = +0.06596$ eV atom$^{-1}$<br><br>$E_B = -6.35589$ eV atom$^{-1}$ |
| $ReN_3$ | 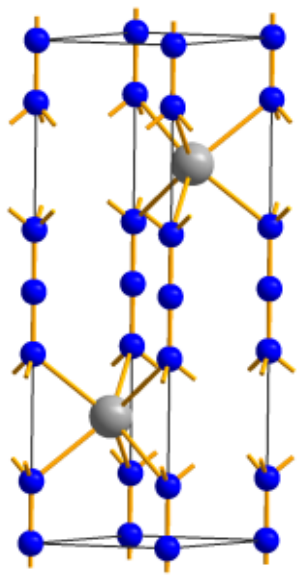 | Space Group: $P6_3/mmc$ (194)<br><br>$a_0 = b_0 = 2.90$ Å,<br>$c_0 = 10.41$ Å<br>$\alpha = \beta = 90$; $\gamma = 120$<br><br>Atomic Positions:<br>Re Wyckoff 2*d*<br>(0.66667, 0.33333, 0.25)<br>N Wyckoff 4*e*<br>(0, 0, 0.12365)<br>N Wyckoff 2*a*<br>(0, 0, 0) | $V_0 = 75.79$ Å$^3$<br><br>$B_0 = 293.6$ GPa<br><br>$dB_0 = 7.09$<br><br>$E_t = -67528.706169$ Ry<br><br>$E_A = +0.64731$ eV atom$^{-1}$<br><br>$E_B = -6.57727$ eV atom$^{-1}$ |
| $ReN_3$ | 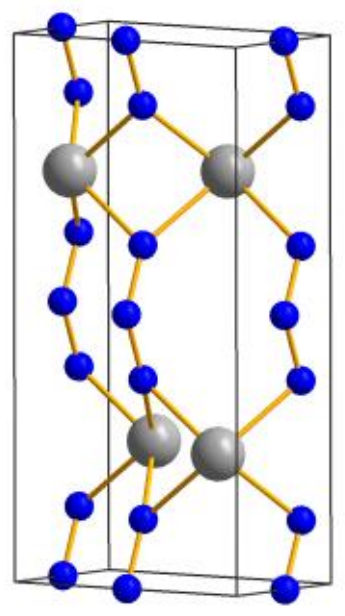 | Space Group: $Am2a$ (40)<br><br>$a_0 = 10.06$ Å, $b_0 = 5.00$ Å,<br>$c_0 = 2.95$ Å<br>$\alpha = \beta = \gamma = 90$<br><br>Atomic Positions:<br>Re Wyckoff 4*b*<br>(0.25, 0.66751, 0.30367)<br>N Wyckoff 8*c*<br>(0.12255, 0.00012, 0.31473)<br>N Wyckoff 4*a*<br>(0, 0, 0.48985) | $V_0 = 148.42$ Å$^3$<br><br>$B_0 = 321.9$ GPa<br><br>$dB_0 = 4.7897$<br><br>$E_t = -67528.791356$ Ry<br><br>$E_A = +0.50243$ eV atom$^{-1}$<br><br>$E_B = -6.72215$ eV atom$^{-1}$ |

[1] $E_t(Re_{HCP} = -31252.13779$ Ry); $E_t(N_2 = -219.04782$ Ry); $E_t(N_{Atomic} = -108.81591$ Ry), values obtained by DFT using the same level of theory than the $ReN_x$ compounds.